\documentclass[preprint,eqsecnum,aps,nofootinbib]{revtex4}
\usepackage[dvips]{graphicx}
\usepackage{amssymb}
\usepackage{amsmath}
\usepackage{amsmath, amsthm, amssymb, mathrsfs}
\usepackage{pstricks}

\begin{document}

\title{Amplitude Damping for Single-Qubit System with Single-Qubit
Mixed-State Environment}  
\author{Eylee Jung$^{1}$, Mi-Ra Hwang$^{1}$, You Hwan Ju$^{1}$, 
               D. K. Park$^{1}$, \\ 
       Hungsoo Kim$^{2}$,
        Min-Soo Kim$^{3}$, Jin-Woo Son$^{3}$}

\affiliation{$^1$ Department of Physics, Kyungnam University, Masan, 631-701, 
Korea    \\
$^2$ The Institute of Basic Science, Kyungnam University, 
                          Masan, 631-70, Korea \\
$^3$ Department of Mathematics, Kyungnam University, Masan, 631-701, Korea}

\begin{abstract}
We study a generalized amplitude damping channel when environment is 
initially in the single-qubit mixed state. Representing the affine
transformation of the generalized amplitude damping by a three-dimensional
volume, we plot explicitly the volume occupied by the channels simulatable
by a single-qubit mixed-state environment. As expected, this volume is 
embedded in the total volume by the channels which is simulated by 
two-qubit enviroment. The volume ratio is approximately $0.08$ which is 
much smaller than $3/8$,
the volume ratio for generalized depolarizing channels.
\end{abstract}


\maketitle

\newpage
\section{Introduction}
About three decades ago R. P. Feynman\cite{feynman82,feynman86} suggested that
a mathematical computation can be efficiently performed by making use of 
quantum mechanics. This suggestion seems to be a starting point for the current
active research of quantum computer. Ten years later after Feynman's suggestion
P. W. Shor\cite{shor94} developed the efficient factoring algorithm for the
large integer in the 
quantum computer. Shor's factoring algorithm makes the most current 
cryptographic methods useless, when the quantum computer is constructed.
Subsequently, the efficient search algorithm was developed
by L. K. Grover\cite{grover96,grover97}. The factoring and search algorithms
were reviewed in Ref.\cite{lavor03} from the physically-motivated aspect.
Recently, Shor's factoring algorithm was realized in NMR\cite{vander01}
and optical\cite{lu07} experiments. In addition, the quantum search algorithm 
was also physically realized in Ref.\cite{kwiat00,walther05}

The quantum computer uses frequently the unitary evolution of the closed 
quantum system. If, however, the quantum system interacts with environment,
the system takes the unwanted non-unitary evolution, which appears as noise
in quantum information processing. Therefore we should understand
and control such noise process\cite{nielsen00}.

In this paper we would like to study on the effect of the environment when
the principal system is a single-qubit pure state. 
In order for the principal system to evolve generally 
it is well-known that 
we need  two-qubit environment\cite{schum96}. However, 
Ref.\cite{lloyd96} argued that one-qubit mixed-state environment might be
sufficient to simulate the most general quantum evolution of a single-qubit
system. Ref.\cite{lloyd96} conjectured this argument by counting the numbers
of independent parameters. 

Later, however, many single-qubit principal channels were found, which cannot 
be simulated by single-qubit environment\cite{terhal99,bacon01}.
Furthermore, recently, Ref.\cite{narang07} has shown that only $3/8$ of the 
generalized depolarizing channels can be simulated by the one-qubit mixed-state
environment. 

In this paper we would like to extend Ref.\cite{narang07} by examing 
the amplitude damping channel. The amplitude damping is an important quantum noise, 
which describes the effect of energy dissipation. The quantum noise is usually
explored using a quantum operation $\varepsilon (\rho)$, which is a {\it convex-linear}
map from density operator of the input space to that of output space, i.e.
$\rho_{out} = \varepsilon (\rho_{in})$\cite{nielsen00}. In this langusge the amplitude
damping is described via operator-sum representation as 
\begin{equation}
\label{add1}
\varepsilon_{AD} (\rho) = E_0 \rho E_0^{\dagger} + E_1 \rho E_1^{\dagger}
\end{equation}
where operation elements $E_0$ and $E_1$ are 
\begin{eqnarray}
\label{add2}
E_0 = \left(    \begin{array}{cc}
               1      &      0  \\
               0     &     \sqrt{1 - \gamma}
                 \end{array}                \right)   \hspace{1.0cm}
E_1 = \left(     \begin{array}{cc}
               0      &     \sqrt{\gamma}   \\
               0      &      0
                 \end{array}                 \right)
\end{eqnarray}
and the parameter $\gamma$ represents the probability for energy loss due to 
losing a particle. Since the density operator of single qubit system can be 
always expressed as $\rho = (\openone + \vec{r} \cdot \vec{\sigma}) / 2$ and
$\varepsilon (\rho) = (\openone + \vec{s} \cdot \vec{\sigma}) / 2$  where
$\sigma_i$'s are Pauli matrices, the amplitude damping (\ref{add1}) can be 
differently expressed from one Bloch vector $\vec{r}$ to another Bolch vector
$\vec{s}$ in the following:
\begin{eqnarray}
\label{add3}
\left(  \begin{array}{c}
             s_1    \\
             s_2    \\
             s_3
        \end{array}     \right) =
\left(   \begin{array}{ccc}
         \sqrt{1 - \gamma} & 0 & 0  \\
         0  &  \sqrt{1 - \gamma} & 0  \\
         0  &  0  &  1 - \frac{\gamma}{2}
          \end{array}            \right)
\left(    \begin{array}{c}
             r_1    \\
             r_2    \\
             r_3
        \end{array}     \right)   + 
\left(     \begin{array}{c}
                0    \\
                0    \\
              \frac{\gamma}{2}
           \end{array}          \right).
\end{eqnarray}
The map from
$\vec{r}$ to $\vec{s}$ is called affine map and it, in general, is very useful to 
visulize the effect of quantum operation in Bloch sphere. In this paper we will 
generalize the amplitude damping and its corresponding affine map. Making use of the
generalized map we will plot explicitly the three-dimensional volume, each point 
inside of which represents a state which 
can be reached from pure initial state when the environment is two-qubit pure state. 
The volume is
compared with another volume derived from the single-qubit mixed-state environment. It
will be shown graphically that the latter volume is embedded in the former, which 
indicates that the single-qubit mixed-state environment cannot simulate the whole
channels derived from two-qubit environment.

This paper is organized as follows.
In Sec. II we briefly review Ref.\cite{narang07}. In Sec. III the 
generalized amplitude damping(GAD) is considered. It is shown that the affine
map of GAD allows the double-degenerate transformation matrix $M$.  
It also allows that only the last component of the translation vector
$\vec{C}$ is nonvanishing.
In Sec. IV we tried to find the GAD when the environment is single-qubit
mixed state. It is shown by plotting the three-dimensional volume that the
GAD channels simulated from  the single-qubit environment have 
very small portion compared
to those simulated from two-qubit environment. 
The volume ratio is numerically computed and is approximately $0.08$, 
which is much smaller than the ratio 
$3/8$ for generalized depolarizing channels.
Sec. V summarizes conclusion and further research direction briefly..

\section{Brief Review: One-qubit system with one qubit environment}
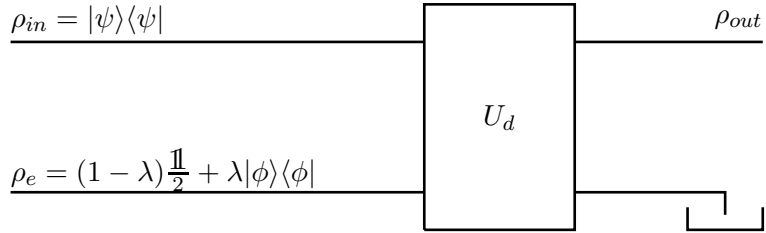
\begin{figure}
\begin{center}
\begin{pspicture}(-7,-2)(4,2)
   \psline[linewidth=1pt](-6.5, 1.0)(-1.0, 1.0)
   \psline[linewidth=1pt](-6.5,-1.0)(-1.0,-1.0)
   \rput[l](-6.5, 1.3){$ \rho_{in} = |\psi\rangle \langle\psi| $}
   \rput[l](-6.5,-0.7){$ \rho_{e} = (1-\lambda) \frac{\openone}{2} 
    + \lambda |\phi\rangle 
                                                                 \langle\phi| $}
   
   \psline[linewidth=1pt](-1,-1.5)(-1,1.5)(1,1.5)(1,-1.5)(-1,-1.5)
   
   \rput(-0.0, 0){$\Large{U_d}$}
   
   \rput[r](3.5,1.3){$ \rho_{out}$}
   \psline[linewidth=1pt](1, 1.0)(3.5, 1.0)
   \psline[linewidth=1pt](1,-1.0)(3.0,-1.0)(3.0,-1.3)
   \psline[linewidth=1pt]( 2.5,-1.2)(2.5,-1.5)(3.5,-1.5)(3.5,-1.2)

\end{pspicture}
\end{center}
\caption{Circuit model for the single-qubit channel in the presence of the
single-qubit mixed-state environment. The principal system is in $|\psi\rangle$ 
initially and the environment is in a mixed state $\rho_e$. After unitary
interaction via $U_d$, the environment will be traced out.}
\end{figure}

In this section we consider a composed closed system which consists of 
one-qubit principal system and one-qubit mixed-state environment as 
pictorially depicted in Fig. 1. Since similar situation was rigorously 
discussed elsewhere\cite{narang07}, we would like to review it briefly. 

We assume the principal system is initially in the pure state,
{\it i.e.} $\rho_{in} = |\psi\rangle \langle\psi |$, where
\begin{equation}
\label{ini-sys}
|\psi\rangle = \cos \frac{\theta}{2} |0\rangle + e^{-i \phi} \sin \frac{\theta}{2} 
                               |1\rangle.
\end{equation}
This state is represented as a point in the Bloch sphere\cite{nielsen00}.

Next we define the initial state of the environment. In order to control
the mixed status of the initial state we introduce a real parameter $\lambda$
and define 
\begin{equation}
\label{ini-env}
\rho_e = (1 - \lambda) \frac{\openone}{2} + \lambda |\phi\rangle \langle\phi|
\end{equation}
where 
\begin{equation}
\label{defphi}
|\phi\rangle =  \cos \frac{\xi}{2} |0\rangle + e^{-i \eta} \sin \frac{\xi}{2} |1\rangle.
\end{equation}
Thus $\lambda = 0$ and $\lambda = 1$ correspond to the completely mixed state
and pure state, respectively. If $0 < \lambda < 1$, the environment is in the 
partially mixed state.

Since the joint system is assumed to be closed, the interaction between the 
physical system and the environment is represented by the unitary 
matrix $U_d$, which is an element of $SU(4)$. Thus this evolution matrix
has generally fifteen free parameters. 
As Ref.\cite{kraus01} has shown, however, the number of these free parameters 
can be reduced to three by making use of the local $SU(2)$ unitary operators.
Furthermore, it was shown in the same reference that this three-parameter
family of $U_d$ is simply expressed in the Bell basis. 
Transforming the matrix representation of $U_d$ into the computational basis
with discarding the unimportant global phase factor simply yields
\begin{eqnarray}
\label{unitary2}
U_d = \left(     \begin{array}{cccc}
\cos \frac{\alpha + \gamma}{2}  &  0  &  0  & i \sin \frac{\alpha + \gamma}{2}
                                                                       \\
0  &  \cos \frac{\alpha - \gamma}{2} e^{-i \beta}  & 
    i \sin \frac{\alpha - \gamma}{2} e^{-i \beta}  &  0                \\
0  &  i \sin \frac{\alpha - \gamma}{2} e^{-i \beta} & 
      \cos \frac{\alpha - \gamma}{2} e^{-i \beta}  &  0                \\
i \sin \frac{\alpha + \gamma}{2}  &  0  &  0  &  \cos \frac{\alpha + \gamma}{2}
                  \end{array}
                                              \right)
\end{eqnarray}
where $\alpha$, $\beta$ and $\gamma$ are real free parameters.

Since $\rho_{in}$, $\rho_e$, and $U_d$ are given, $\rho_{out}$ can be 
explicitly computed by unitary evolution and partial trace in the following
\begin{equation}
\label{rhoout1}
\rho_{out} = \mbox{Tr}_{env} \left[ U_d (\rho_{in} \otimes \rho_e) U_d^{\dagger}
                                        \right].
\end{equation}
Let us assume $\rho_{in} = (\openone + \vec{r} \cdot \vec{\sigma} ) / 2$ and 
$\rho_{out} = (\openone + \vec{r}' \cdot \vec{\sigma} ) / 2$.
Then the quantum operation defined 
\begin{equation}
\label{operation1}
\varepsilon (\rho_{in}) = \rho_{out}
\end{equation}
is given by the affine map
\begin{equation}
\label{affine1}
r_i \rightarrow r'_i = M_{i j} r_j + C_j
\end{equation}
where $M_{i j}$ is $3 \times 3$ real matrix in the form
\begin{eqnarray}
\label{affine2}
M_{i j} = \left(   \begin{array}{ccc}
\cos \beta \cos \gamma  &  \lambda \cos \xi \sin \beta \cos \gamma
                        &  -\lambda \sin \xi \sin \eta \cos \beta \sin \gamma
                                                          \\
-\lambda \cos \xi \cos \alpha \sin \beta  &  \cos \alpha \cos \beta &
\lambda \sin \xi \cos \eta \sin \alpha \cos \beta
                                                          \\
\lambda \sin \xi \sin \eta \cos \alpha \sin \gamma  &
-\lambda \sin \xi \cos \eta \sin \alpha \cos \gamma  &
\cos \alpha \cos \gamma
                     \end{array}          \right)
\end{eqnarray}
and the column vector $\vec{C}$ is 
\begin{eqnarray}
\label{affine3}
\vec{C} = -\lambda \left(   \begin{array}{c}
   \sin \xi \cos \eta \sin \beta \sin \gamma \\
   \sin \xi \sin \eta \sin \alpha \sin \beta  \\
   \cos \xi \sin \alpha \sin \gamma
                             \end{array}   \right).
\end{eqnarray}
This affine map gives a parametrization of all the channels simulated by a 
one-qubit mixed-state environment. Varying the six parameters $\alpha$, 
$\beta$, $\gamma$, $\lambda$, $\xi$, and $\eta$, we can obtain the various
output states $\rho_{out}$. We can use this various output states to explore 
the damping effect of the principal system arising due to the interaction with
environment. 

For later use we would like to discuss the eigenvalues $\Lambda$ of 
$M^{\dagger} M$.  To compute $\Lambda$ we should solve the highly complicated
third-order equation
\begin{equation}
\label{eigen1}
-\Lambda^3 + f_1 \Lambda^2 + f_2 \Lambda + f_3 = 0
\end{equation}
where
\begin{eqnarray}
\label{eigen2}
& &f_1 = \cos^2 \alpha \cos^2 \beta + \cos^2 \beta \cos^2 \gamma + 
      \cos^2 \gamma \cos^2 \alpha 
                                       \\   \nonumber
& & + \lambda^2 \Bigg[ \cos^2 \xi  \sin^2 \beta 
      (\cos^2 \alpha + \cos^2 \gamma) + \sin^2 \xi \sin^2 \eta \sin^2 \gamma
      (\cos^2 \alpha + \cos^2 \beta) 
                                        \\   \nonumber
& & \hspace{4.0cm} + \sin^2 \xi \cos^2 \eta \sin^2 \alpha
      (\cos^2 \beta + \cos^2 \gamma) \Bigg]   \\  \nonumber
& &f_2 = -\Bigg[\cos^2 \alpha \cos^2 \beta \cos^2 \gamma + \lambda^2 \bigg(
             \sin^2 \xi \sin^2 \eta \cos^2 \alpha \cos^2 \beta \sin^2 \gamma 
                                                       \\  \nonumber
& &\hspace{2.0cm} +
             \sin^2 \xi \cos^2 \eta \sin^2 \alpha \cos^2 \beta \cos^2 \gamma +
             \cos^2 \xi \cos^2 \alpha \sin^2 \beta \cos^2 \gamma 
                                                 \bigg) \Bigg]
                                                       \\  \nonumber
& &\times \Bigg[(\cos^2 \alpha + \cos^2 \beta + \cos^2 \gamma) + 
      \lambda^2 \bigg( \sin^2 \xi \sin^2 \eta \sin^2 \gamma +
                       \sin^2 \xi \cos^2 \eta  \sin^2 \alpha + 
                       \cos^2 \xi \sin^2 \beta \bigg) \Bigg]
                                                        \\  \nonumber
& &f_3 = \Bigg[\cos^2 \alpha \cos^2 \beta \cos^2 \gamma + 
              \lambda^2 \bigg( \cos^2 \xi \cos^2 \alpha \sin^2 \beta 
                                                          \cos^2 \gamma +
                \sin^2 \xi \sin^2 \eta \cos^2 \alpha \cos^2 \beta \sin^2 \gamma
                                                        \\   \nonumber
& & \hspace{7.0cm}
             + \sin^2 \xi \cos^2 \eta \sin^2 \alpha \cos^2 \beta \cos^2 \gamma
                               \bigg)   \Bigg]^2.
\end{eqnarray}
Although one can solve $\Lambda$ analytically in principle, it would be too
lengthy to express them explicitly. When, however, $\alpha = \gamma$, the 
eigenvalues reduce to the simpler expression in the following:
\begin{eqnarray}
\label{eigen3}
& &\Lambda_1 = \cos^2 \alpha \cos^2 \beta + \lambda^2
\left[ \cos^2 \xi \cos^2 \alpha \sin^2 \beta + \sin^2 \xi \sin^2 \alpha
                                               \cos^2 \beta \right]
                                                         \\  \nonumber
& &\Lambda_{\pm} = \frac{\cos^2 \alpha}{2} 
             \left[ (\cos^2 \alpha + \cos^2 \beta) + 
                    \lambda^2 (\cos^2 \xi \sin^2 \beta + 
                         \sin^2 \xi \sin^2 \alpha) \pm \tilde{\Lambda} \right]
\end{eqnarray}
where
\begin{equation}
\label{eigen4}
\tilde{\Lambda} = 
\sqrt{\left\{ (\cos^2 \alpha - \cos^2 \beta) + \lambda^2 
(\cos^2 \xi \sin^2 \beta + \sin^2 \xi \sin^2 \alpha ) \right\}^2 - 4 \lambda^2
\cos^2 \xi \sin^2 \beta (\cos^2 \alpha - \cos^2 \beta )}.
\end{equation}
Another special case is $\xi = 0$, which gives
\begin{eqnarray}
\label{eigen5}
& &\Lambda_1 = \cos^2 \alpha (\cos^2 \beta + \lambda^2 \sin^2 \beta)
                                                           \\  \nonumber
& &\Lambda_2 = \cos^2 \gamma (\cos^2 \beta + \lambda^2 \sin^2 \beta)
                                                           \\   \nonumber
& &\Lambda_3 = \cos^2 \alpha \cos^2 \gamma.
\end{eqnarray}
These eigenvalues will be used to analyze the amplitude damping channels 
simulated by the single-qubit environment.

\section{Generalized Amplitude Damping}
Amplitude damping is a description of energy dissipation -effects due to loss
of energy from a quantum system. The operator-sum representation for the 
amplitude damping channel defined in Eq.(\ref{add1}) can be generalized by
\begin{equation}
\label{ad1}
\rho \rightarrow \rho' = \sum_{i=0}^3 E_i \rho E_i^{\dagger}
\end{equation}
where the operation elements are 
\begin{eqnarray}
\label{ad2}
E_0 = \sqrt{\epsilon_0}
\left(    \begin{array}{cc}
          1   &   0   \\
          0   &  \sqrt{\gamma_0}
           \end{array}             \right)
\hspace{2.0cm}
E_1 = \sqrt{\epsilon_1}
\left(    \begin{array}{cc}
          0   &   \sqrt{\gamma_1}   \\
          0   &  0
           \end{array}             \right)
                                              \\  \nonumber
E_2 = \sqrt{\epsilon_2}
\left(    \begin{array}{cc}
          \sqrt{\gamma_2}   &   0   \\
          0   &                   1
           \end{array}             \right)
\hspace{2.0cm}
E_3 = \sqrt{\epsilon_3}
\left(    \begin{array}{cc}
          0   &   0   \\
          \sqrt{\gamma_3}   &  0
           \end{array}             \right)
\end{eqnarray}
with
\begin{equation}
\label{ad3}
\epsilon_0 + \gamma_2 \epsilon_2 + \gamma_3  \epsilon_3
= \gamma_0 \epsilon_0 + \gamma_1 \epsilon_1 + \epsilon_2
= 1.
\end{equation}
The fact $\sum_i E_i^{\dagger} E_i = I$ implies that the quantum operation
for the amplitude damping is a trace-preserving map. Since there are four 
operation elements, the GAD is realized when the
environment is two-qubit system. Therefore, a natural question arises: how 
much portion for the amplitude damping can be simulated when the environment
is a single-qubit mixed state? This question is related to the volume
issue, which will be discussed in the next section.

The amplitude damping defined in Eq.(\ref{ad1}) can be described by the
affine map
\begin{eqnarray}
\label{ad3}
\left( \begin{array}{c}
        \tilde{r}_1   \\
        \tilde{r}_2   \\
        \tilde{r}_3
       \end{array}   \right) = M_{AD}
\left( \begin{array}{c}
        r_1   \\
        r_2   \\
        r_3
       \end{array}   \right) + \vec{C}_{AD}
\end{eqnarray}
where
\begin{eqnarray}
\label{ad4}
& &M_{AD} = 
\left(          \begin{array}{ccc}
\epsilon_0 \sqrt{\gamma_0} + \epsilon_2 \sqrt{\gamma_2}  &  0  &  0     \\
0  & \epsilon_0 \sqrt{\gamma_0} + \epsilon_2 \sqrt{\gamma_2}   &  0     \\
0  &  0  &  -1 + \epsilon_0 (1 + \gamma_0) +  \epsilon_2 (1 + \gamma_2)
                \end{array}          \right)
                                                    \\   \nonumber
& &
\hspace{3.0cm}
\vec{C}_{AD} =    
\left(           \begin{array}{c}
                  0               \\
                  0               \\
                  \epsilon_0 (1 - \gamma_0)  - \epsilon_2 (1 - \gamma_2)
                  \end{array}                       \right).
\end{eqnarray}
Thus the generalized amplitude damping has following two important 
properties: (i) the transformation matrix $M_{AD}$ has two-fold degeneracy in 
the eigenvalues. (ii) the first two components of the translation vector
$\vec{C}_{AD}$ are zero. As shown in Eq.(\ref{add3}) the standard amplitude damping 
has same properties. This is a reason why we define the GAD as Eq..(\ref{ad1}) and 
(\ref{ad2}).

The most general GAD channels simulated from the
two-qubit environment can be represented by the three-dimensional volume
$(X_2, Y_2, Z_2)$ defined
\begin{eqnarray}
\label{ad5}
X_2&=&\epsilon_0 \sqrt{\gamma_0} + \epsilon_2 \sqrt{\gamma_2}
                                                          \\  \nonumber
Y_2&=&-1 + \epsilon_0 (1 + \gamma_0) + \epsilon_2 (1 + \gamma_2)
                                                          \\   \nonumber
Z_2&=&\epsilon_0 (1 - \gamma_0) - \epsilon_2 (1 - \gamma_2).
\end{eqnarray}
This volume is plotted in Fig. 2 transparently to compare with the volume 
derived from the single-qubit mixed-state environment\cite{address}. 
Compared to the 
depolarizing channel, where the tetrahedron volume is derived\cite{narang07},
the volume for the amplitude damping channel is very complicated. Since,
furthermore, $(X_2, Y_2, Z_2)$ depends on the four parameters
$\epsilon_0$, $\epsilon_2$, $\gamma_0$ and $\gamma_2$, it is highly 
difficult to compute the volume exactly. 
The numerical calculation gives the volume approximately $1.67$.
We will show in the next section that the volume derived from the 
single-qubit mixed-state environment is embedded in this volume.

\section{Volume issue}
In this section we want to explore the amplitude damping 
when the environment is a 
single-qubit mixed-state. In order to simulate the amplitude damping, as 
shown in the previous section, the transformation matrix (\ref{affine2})
should have the following two properties: (i) in the singular value 
decomposition $M = U D V$ where  $U$ and $V$ are unitary matrices, the 
diagonal matrix $D$ should have double degeneracy. (ii) the first
two components of the translation 
vector $U^{\dagger} \vec{C}$ should be zero. 

In this paper  we consider the case of $\alpha = \gamma$, where the eigenvalues
of $M^{\dagger} M$ are somewhat simple.
In this case 
Eq.(\ref{eigen3}) and Eq.(\ref{eigen4}) imply that the necessary condition for
the diagonal matrix $D$ to have the double degeneracy is the removal of the 
square root in $\tilde{\Lambda}$. This condition reduces to the following 
four distinct cases: (1) $\xi = 0$, (2) $\xi = \pi / 2$, (3) $\beta = 0$,
(4) $\alpha = \beta = \gamma$. The diagonal components $D_{11}$, $D_{22}$,
and $D_{33}$ for the diagonal matrix $D$ for each case are summarized 
in Table I.

\begin{center}

\begin{tabular}{c|c}  \hline
Cases   \hspace{.5cm} &    Diagonal components     \\ \hline \hline
$\xi = 0$  &  \hspace{.5cm} 
             $D_{11} = D_{22} = \cos \alpha \sqrt{\cos^2 \beta + \lambda^2
                                                   \sin^2 \beta}$
                                                                 \\
           & $D_{33} = \cos^2 \alpha$                       \\  \hline
$\xi = \frac{\pi}{2}$  & $D_{11} = \cos \alpha \sqrt{\cos^2 \alpha + 
                                            \lambda^2 \sin^2 \alpha}$
                                                                    \\
                      & $D_{22} = \cos \beta \sqrt{\cos^2 \alpha + 
                                            \lambda^2 \sin^2 \alpha}$
                                                                    \\
                      & $D_{33} = \cos \alpha \cos \beta$   \\  \hline
$\beta = 0$   &  $D_{11} = \sqrt{\cos^2 \alpha + \lambda^2 \sin^2 \xi \sin^2
                                 \alpha}$
                                                                    \\
              &  $D_{22} = \cos \alpha \sqrt{\cos^2 \alpha + \lambda^2 
                                                \sin^2 \xi \sin^2
                                 \alpha}$
                                                                    \\
              &  $D_{33} = \cos \alpha$        \\   \hline
$\alpha = \beta$  &  $D_{11} = D_{22} = \cos \alpha 
                      \sqrt{\cos^2 \alpha + \lambda^2 \sin^2 \alpha}$
                                                                     \\
                  &  $D_{33} = \cos^2 \alpha$  \\  \hline
\end{tabular}

\vspace{0.1cm}
Table I:The diagonal components of $D$ for each case. The double-degeneracy
occurs for the cases of $\xi = 0$ and $\alpha = \beta$. 
\end{center}
\vspace{1.0cm} 

Table I indicates that the cases $\xi = \pi / 2$ and $\beta = 0$ are excluded as
candidates for the amplitude damping due to no degeneracy. The singular
value decomposition for the remaining candidates are for $\xi = 0$
\begin{eqnarray}
\label{svd1}
& &U = \left(   \begin{array}{ccc}
              0  &  1   &   0   \\
              -1 &  0   &   0   \\
              0  &  0   &   1
                  \end{array}         \right)
                                                \\  \nonumber 
& &D = \left(       \begin{array}{ccc}
\cos \alpha \sqrt{\cos^2 \beta + \lambda^2 \sin^2 \beta}  &  0  &  0  \\
0  &  \cos \alpha \sqrt{\cos^2 \beta + \lambda^2 \sin^2 \beta}  &  0  \\
0  &  0  &  \cos^2 \alpha
                  \end{array}        \right) 
                                                \\   \nonumber
& &V = \frac{1}{\sqrt{\cos^2 \beta + \lambda^2 \sin^2 \beta}}
\left(       \begin{array}{ccc}
\lambda \sin \beta & -\cos \beta & 0   \\
\cos \beta & \lambda \sin \beta  & 0   \\
0  &  0  &  \sqrt{\cos^2 \beta + \lambda^2 \sin^2 \beta}
              \end{array}             \right)  
\end{eqnarray} 
and for $\alpha = \beta$
\begin{eqnarray}
\label{svd2}
& &U = \frac{1}{\sqrt{}}
\left(       \begin{array}{ccc}
\sin \eta \cos \alpha - \lambda \cos \xi \cos \eta \sin \alpha &
\cos \xi \cos \eta \cos \alpha + \lambda \sin \eta \sin \alpha &
\sin \xi \cos \eta \sqrt{}  \\
-(\cos \eta \cos \alpha + \lambda \cos \xi \sin \eta \sin \alpha)  &
\cos \xi \sin \eta \cos \alpha - \lambda \cos \eta \sin \alpha &
\sin \xi \sin \eta \sqrt{} \\
\lambda \sin \xi \sin \alpha & -\sin \xi \cos \alpha &
\cos \xi \sqrt{}
               \end{array}          \right)
                                                     \nonumber  \\
& &D = \left(     \begin{array}{ccc}
\cos \alpha \sqrt{} & 0 & 0      \\
0 & \cos \alpha \sqrt{} & 0      \\
0 & 0 & \cos^2 \alpha
                \end{array}         \right)
                                                     \\  
& &V = \left(   \begin{array}{ccc}
\sin \eta  &  -\cos \eta &  0    \\
\cos \xi \cos \eta &  \cos \xi \sin \eta &  - \sin \xi   \\  
\sin \xi \cos \eta & \sin \xi \sin \eta &  \cos \xi
                  \end{array}         \right)                 \nonumber
\end{eqnarray}
respectively, 
where $\sqrt{} = \sqrt{\cos^2 \alpha + \lambda^2 \sin^2 \alpha}$.
Computing the translation vector $U^{\dagger} \vec{C}$, one can show that the 
amplitude damping derived from the single-qubit mixed state environment is 
represented by the three-dimensional volume $(X_1, Y_1, Z_1)$ defined
\begin{eqnarray}
\label{three1}
& &X_1 \equiv D_{11} = \cos \alpha \sqrt{\cos^2 \beta + \lambda^2 \sin^2 \beta}
                                                    \\  \nonumber
& &Y_1 \equiv D_{33} = \cos^2 \alpha
                                                    \\   \nonumber
& &Z_1 \equiv (U^{\dagger} \vec{C})_3 = -\lambda \sin^2 \alpha.
\end{eqnarray}

\begin{figure}[ht!]
\begin{center}
\includegraphics[height=10cm]{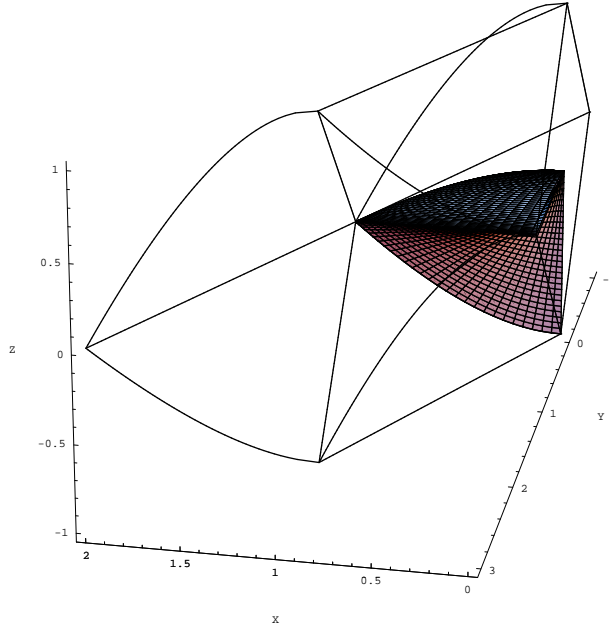}
\caption[fig2]{Graphical representation of the volumes occupied by 
$(X_2, Y_2, Z_2)$ (transparent volume) and $(X_1, Y_1, Z_1)$ (opaque volume).
As expected the opaque volume is embedded into the transparent volume. This 
fact indicates that the amplitude damping channel cannot by simulated 
completely by the one-qubit environment.} 
\end{center}
\end{figure}

The volume generated by $(X_1, Y_1, Z_1)$ is plotted in Fig. 2 
opaquely\cite{address}.
As expected this volume is embedded in the lucid volume generated by 
$(X_2, Y_2, Z_2)$. This means that the amplitude damping channel cannot be
completely simulated by the one-qubit environment although it is in the
arbitrary mixed-state as depolarizing channel. 
The volume for $(X_1, Y_1, Z_1)$ can be computed analytically, which is 
$2/15$. 
Thus the volume ratio, {\it i.e.}
opaque volume divided by transparent volume, is approximately $0.08$. This is 
much smaller than
$3/8$, which is the volume ratio for the depolarizing channel.

\section{conclusion}
We have studied the GAD channels simulated by the one-qubit mixed-state 
environment when the principal system is initially in the single-qubit
pure state. Examing the affine map for the GAD channel simulated by 
two-qubit environment, we have found that $\xi = 0$ with $\alpha = \gamma$,
and $\alpha = \beta = \gamma$ are the GAD channel simulated by the one-qubit
mixed-state environment. Representing the affine map as a three-dimensional 
volume, we have plotted the volume opaquely in Fig. 2. As expected, 
this volume is 
embedded in the total volume generated by the two-qubit environment. It turns 
out that the volume ratio is much smaller than $3/8$, which
is the volume ratio for the depolarizing channel. 

It seems to be interesting to explore the various different damping channels
in this way. For example, let us consider the phase damping whose quantum operation 
is defined as $\varepsilon(\rho) = E_0 \rho E_0^{\dagger} + E_1 \rho E_1^{\dagger}$,
where operation elements are 
\begin{eqnarray}
\label{cadd1}
E_0 = \left(    \begin{array}{cc}
                1    &    0     \\
                0    &   \sqrt{1 - \lambda}
                \end{array}                   \right)   \hspace{2.0cm}
E_1 = \left(    \begin{array}{cc}
                0    &    0     \\
                0    &   \sqrt{\lambda}
                \end{array}                   \right)
\end{eqnarray}
and $\lambda$ is a quantity related to a relaxation time.
The affine map for the phase damping is thus 
$(r_1, r_2, r_3) \rightarrow (r_1 \sqrt{1 - \lambda}, r_2 \sqrt{1 - \lambda}, r_3)$. 
Therefore the effect of the phase damping is to shrink the Bloch sphere into ellipsoid.
To explore the effect of one-qubit mixed-state environment in the phase damping 
process firstly we should generalize it by introducing four operation elements with
keeping the general features of the standard phase damping. Next we should find 
same channels when the environment is single-qubit mixed states with making use
of Eq.(\ref{affine2}). It is unclear at least for us how to construct the generalized
phase damping. 

Another direction we would like to explore is to compute the entanglement measure
when the environment is involved. Recently, the Groverian measure for mixed 
states was introduced in Ref.\cite{shapira06}
Although it was shown in Ref.\cite{shapira06} that the Groverian measure for mixed
states is entanglement monotone, the explicit computation of it for given mixed 
states is highly nontrivial mainly due to the maximization over purification while 
the analytic computation for the pure states is sometimes 
possible\cite{tama07-1}. Since 
environment in general makes the state of quantum system mixed state, it seems to be
highly interesting to explore the role of entanglement in the damping process.

\vspace{1cm}

{\bf Acknowledgement}:  
This work was supported by the Kyungnam University
Research Fund, 2006.

\end{document}